\documentclass[prb,floatfix,preprint,superscriptaddress]{revtex4-1} 

\usepackage[amssymb]{SIunits}

\pdfoutput=1 
                                                                                          
\usepackage[bookmarks=false]{hyperref} 
\usepackage{graphicx}
\usepackage{textcomp}
\usepackage[english]{babel}
\usepackage[utf8]{inputenc}
\usepackage[T1]{fontenc}
\usepackage{hyperref}
\usepackage{amsmath}
\usepackage{amssymb}
\usepackage[amssymb]{SIunits}

\newcommand{\UGAAddress}{Univ. Grenoble Alpes, F-38000 Grenoble, France}
\newcommand{\CNRSAddress}{CNRS, Inst. NEEL, "Nanophysique et semiconducteurs" group, F-38000 Grenoble, France}
\newcommand{\CEAAddress}{CEA, INAC, "Nanophysique et semiconducteurs" group, F-38000 Grenoble, France}
\newcommand{\LETIAddress}{CEA, LETI, MINATEC Campus, F-38054 Grenoble, France}

\begin{document}
\title{Cathodoluminescence spectroscopy of plasmonic patch antennas: towards lower order 
and higher energies}

\author{Mathieu Jeannin}
\affiliation{\UGAAddress}
\affiliation{\CNRSAddress}

\author{Névine Rochat}
\affiliation{\UGAAddress}
\affiliation{\LETIAddress}

\author{Kuntheak Kheng}
\affiliation{\UGAAddress}
\affiliation{\CEAAddress}

\author{Gilles Nogues}
\affiliation{\UGAAddress}
\affiliation{\CNRSAddress}

%
%


\begin{abstract}
	We report on the cathodoluminescence characterization of Au, Al and a Au/Al bimetal circular 
	plasmonic patch antennas, with disk diameter ranging from 150 to \unit{900}{\nano\meter}. It allows us
	access to monomode operation of the antennas down to the fundamental dipolar mode, in contrast to 
	previous studies on similar systems. Moreover we show that we can can shift the 
	operation range of the antennas towards the blue spectral range by using Al. Our experimental results
	 are compared to a semi-analytical model that provides qualitative insight on the mode 
	structure sustained by the antennas.
\end{abstract}


\maketitle


\section{Introduction}

Advances on classical and quantum solid state light emitters and optoelectronic devices
bring great interests in controlling light emission properties of nano-sized emitters. 
As an example, several approaches are
employed in the field of semiconducting quantum dots, from microcavities
\cite{Gerard1998} and other photonic structures
\cite{Claudon2010,Lodahl2015,Gschrey2015} to plasmonic antennas
\cite{Curto2010,Pfeiffer2010,Curto2013,Jeannin2016}. In the latter case, a great
effort has been made to develop new antenna geometries that could increase the
coupling strength between a single emitter and the antenna. Additionally,
researchers are investigating novel possibilities of controlling the radiation
pattern of the coupled structure. One of the emerging strategies is to confine
the plasmon field inside an insulating layer comprised between a metallic
nanoparticle and a continuous metallic film. Such structures were initially
investigated using colloidal plasmonic particles \cite{Mock2008,Mubeen2012}. Due
to the strong electric field enhancement possible inside the insulating layer,
these systems were soon proposed as promising plasmonic cavities. Recent
demonstrations include a 1900-fold increase in emission intensity for colloidal
quantum dots \cite{Hoang2016}, and reaching up to the strong coupling regime for
single molecules placed between colloidal Au spheres and an Au mirror
\cite{Chikkaraddy2016}. All of these systems rely on the coupling between the
dipolar localized surface plasmon mode supported by the nanoparticle and the
metallic mirror. Another key properties of the particle-on-a-mirror geometry is
that the coupling between the plasmon mode supported by the particle and the
film allows to dramatically change the emission diagram of the coupled system.
This was especially investigated using Au disk antennas on a Au mirror. Because
of geometrical and electric field profile similarities to their radio-frequency
counterparts, these laterally confined insulator-metal-insulator-metal (IMIM)
nanostructures were then designated as \emph{plasmonic patch antennas}
\cite{Esteban2010,Filter2012,Bigourdan2014}. They have been successfully coupled
to colloidal quantum dots \cite{Belacel2013} and studied by cathodoluminescence
\cite{Mohtashami2014}. Contrary to the nanoparticle-on-a-mirror experiments,
circular patch antennas reported up to now have large diameters. They can be
thought as circular cavities for surface plasmon polariton modes. Their large
size and subsequent large number of supported mode is the key to the control of
their radiation pattern.

In this article, we use cathodoluminescence (CL) spectroscopy and imaging on Au
and Al circular plasmonic patch antennas in the intermediate regime between
large antenna and small particle. We characterize circular antennas supporting
single mode operation as well as highly multipolar modes. The CL signal
collected from plasmonic antennas is closely related to the electromagnetic
local density of states (LDOS). In~\cite{GarciadeAbajo2010, Mohtashami2014,Losquin2015}, 
it is shown that the signal measured in CL corresponds to the integration over the electron 
beam path of the partial, radiative LDOS projected along the electron beam path. 
It allows to precisely characterize the antennas, to determine their spectral
properties and to map the radiative LDOS on a nanometer scale
for each antenna resonance. 
The paper is organized as follows: we first present
the patch antennas fabrication and the CL system. We then introduce the
fundamental properties of Au patch antennas and a semianalytical model proposed
to explain their mode structure \cite{Filter2012, Mohtashami2014}. We then study
antennas of smaller dimensions, down to $\sim$\unit{170}{\nano\meter} in
diameter, allowing for the study of the fundamental antenna mode. Finally, using
Al patch antennas we demonstrate that changing the antenna material makes it
possible to shift their operation range down to a wavelength of
\unit{450}{\nano\meter}. Indeed, Al is seen as a promising candidate for blue
and U.V. plasmonics \cite{Knight2012,Gerard2014,Martin2014,Sobhani2015} in spite
of the current low quality of Al films \cite{McPeak2015}. It has the additional
advantage to be compatible with current CMOS fabrication technologies, easing a
possible large-scale implementation of Al-based plasmonic devices.

\section{Sample fabrication and experimental details}

\begin{figure}[htbp]
\centering\includegraphics[width=\linewidth]{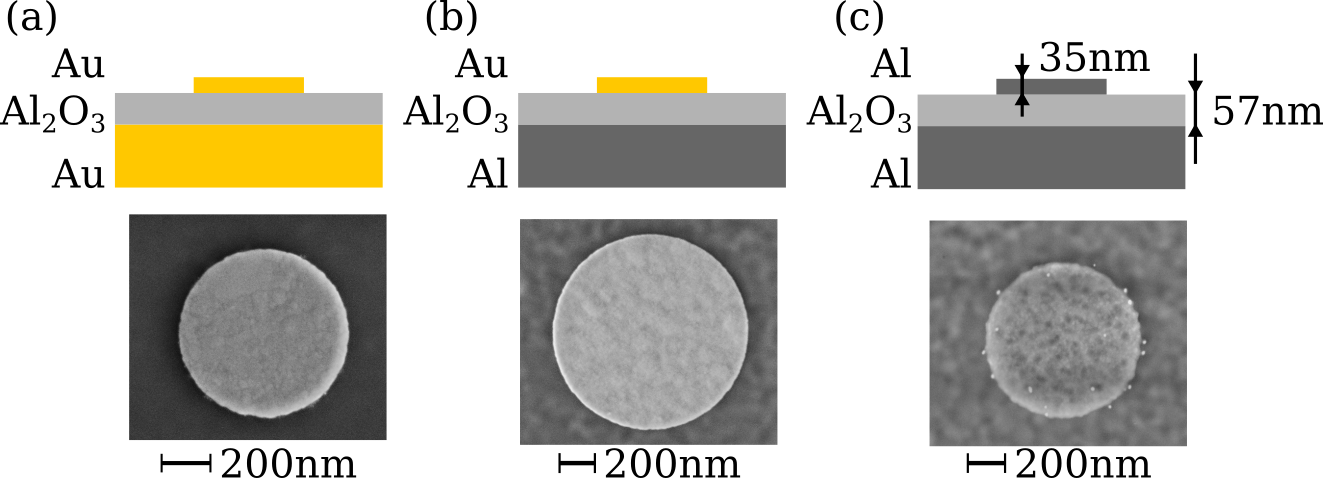}
\caption{Sketch of the three patch antennas configurations: (a) Au patch and Au mirror, (b) Au patch 
and Al mirror, and (c) Al patch and Al mirror. In each case, the Al$_2$O$_3$ spacing layer thickness 
is \unit{57}{\nano\meter}. Below each sketch, we show an SEM image of a typical antenna. Notice the 
difference in surface roughness from (a) to (c) due to the poor quality of the thick Al film. }
\label{fig:PatchsPresentation}
\end{figure}

The patch antennas are fabricated on Si substrates. A first, optically thick
layer of metal (Au or Al, $\unit{100}{\nano\meter}$) is evaporated on the
substrate using e-beam evaporation. In order to get the best optical properties,
we aimed at lowering as much as possible the surface roughness and granular
structure of the films. Au evaporation is performed under $\sim
\unit{10^{-6}}{\milli\bbar}$ at an evaporation rate of $\sim
\unit{2}{\angstrom\per\second}$. Al has a tendency to oxidise a lot more easily
than Au, and therefore smooth Al films are hard to obtain with the vacuums
levels available in conventional e-beam evaporation machines. A possible way to
improve the quality of the film is to use a higher evaporation rate, in our case
$\unit{2}{\nano\meter\per\second}$ \cite{McPeak2015}. The spacing oxide layer is
then deposited using atomic layer deposition, and its thickness is controlled
using an interferometric measurement after deposition.

All the top layer plasmonic structures are fabricated by electron beam
lithography. An electron-sensitive, positive resist (Poly(metyl methacrylate),
or PMMA) is spin-coated on the sample and soft-baked on a hot plate at
\unit{180}{\celsius} for \unit{5}{\minute}. The disks are patterned using an electron
beam exposing the resist upon electron impact. Each disk is separated from its
neighbours by a distance greater than \unit{2}{\micro\meter} to avoid any
coupling between two adjacent structures. The resist is then developed, and
\unit{35}{\nano\meter} of metal (Au or Al) is deposited using electron gun
evaporation. A final lift-off process is performed in 1-methyl-2-pyrrolidinone
(NMP) heated at \unit{80}{\celsius} to remove the remaining resist.

Three IMIM systems are under investigation. In each case, the Al$_2$O$_3$ oxide layer
thickness is 57$\pm$\unit{2}{\nano\meter}. First, Au patch antennas were fabricated on
an Au mirror, as shown in Fig.~\ref{fig:PatchsPresentation}(a), with diameters
ranging from 280 to \unit{900}{\nano\meter}. Second, Au patch antennas on an Al
mirror were fabricated with diameters ranging from 120 to
\unit{930}{\nano\meter} (Fig.~\ref{fig:PatchsPresentation}(b)). Finally, Al
antennas on an Al mirror were fabricated, with diameters also ranging from
120 to \unit{930}{\nano\meter} (Fig.~\ref{fig:PatchsPresentation}(c)). As can be
seen in the scanning electron microscope (SEM) images in
Fig.~\ref{fig:PatchsPresentation}, the Au deposition results in a very smooth Au
film and antenna surface. Magnified SEM images revealed an Au grain size of
around \unit{10}{\nano\meter}. The Al film appears a lot rougher, as can be
inferred from Fig.~\ref{fig:PatchsPresentation}(c). We noticed that Al grains
tend to coalesce, so the smoothness of the top surface degrades upon increasing
the deposited thickness. Hence, the top surface of the Al mirror
(\unit{100}{\nano\meter} thick) shows Al grains sizes of around
\unit{40}{\nano\meter}, and is a lot rougher than the Al antenna surface
(\unit{35}{\nano\meter} thick) which has Al grains of only
\unit{20}{\nano\meter} in size. Note that in the case of bimetal antennas the
roughness of the Al film also slightly degrades the quality of the Au top disk,
as seen in Fig.~\ref{fig:PatchsPresentation}(b).

Our cathodoluminescence setup consists in FEI (quanta 200) SEM fitted with a
drilled home made asymmetric parabolic mirror allowing for electrons to pass
through, and providing a very large numerical aperture (NA) ranging from 0.5
(detector side) to 0.9 (parabola side). We use an acceleration voltage of
\unit{30}{\kilo\electronvolt} and a beam current around \unit{7}{\nano\ampere}.
The hole allowing for the electrons to pass is \unit{500}{\micro\meter} in
diameter, preventing light collection in the vertical direction with an angular
divergence of a few degrees. The CL signal collected by the mirror is focused on
the entrance slit of a spectrometer (Horiba Jobin-Yvon IHR550) by a spherical
mirror, preventing chromatic aberrations. The light is then dispersed by a
150gr/mm grating blazed at \unit{550}{\nano\meter} and sent either to a
charged-coupled device (CCD) camera (Andor Newton) or energy-filtered through
the exit slit of the spectrometer and sent to an avalanche photodiode (APD) to
obtain energy-selective CL images.

Complementary experiments were carried out using a Attolight commercial Rosa CL
setup allowing for hyperspectral imaging. The electron beam acceleration voltage
is \unit{10}{\kilo\electronvolt}, with a beam current of
\unit{25}{\nano\ampere}. The light is collected using a Cassegrain-type
objective (NA 0.72) embedded inside the electronic column of the scanning
electron microscope. The collected light is focused on the entrance slit of a
spectrometer (Horiba Jobin-Yvon IHR320), dispersed using a 150gr/mm grating
blazed at \unit{500}{\nano\meter}, and detected using a CCD camera (Andor
Newton).

\section{Cathodoluminescence of Au patch antennas}

\begin{figure}[htbp]
\centering\includegraphics[width=\linewidth]{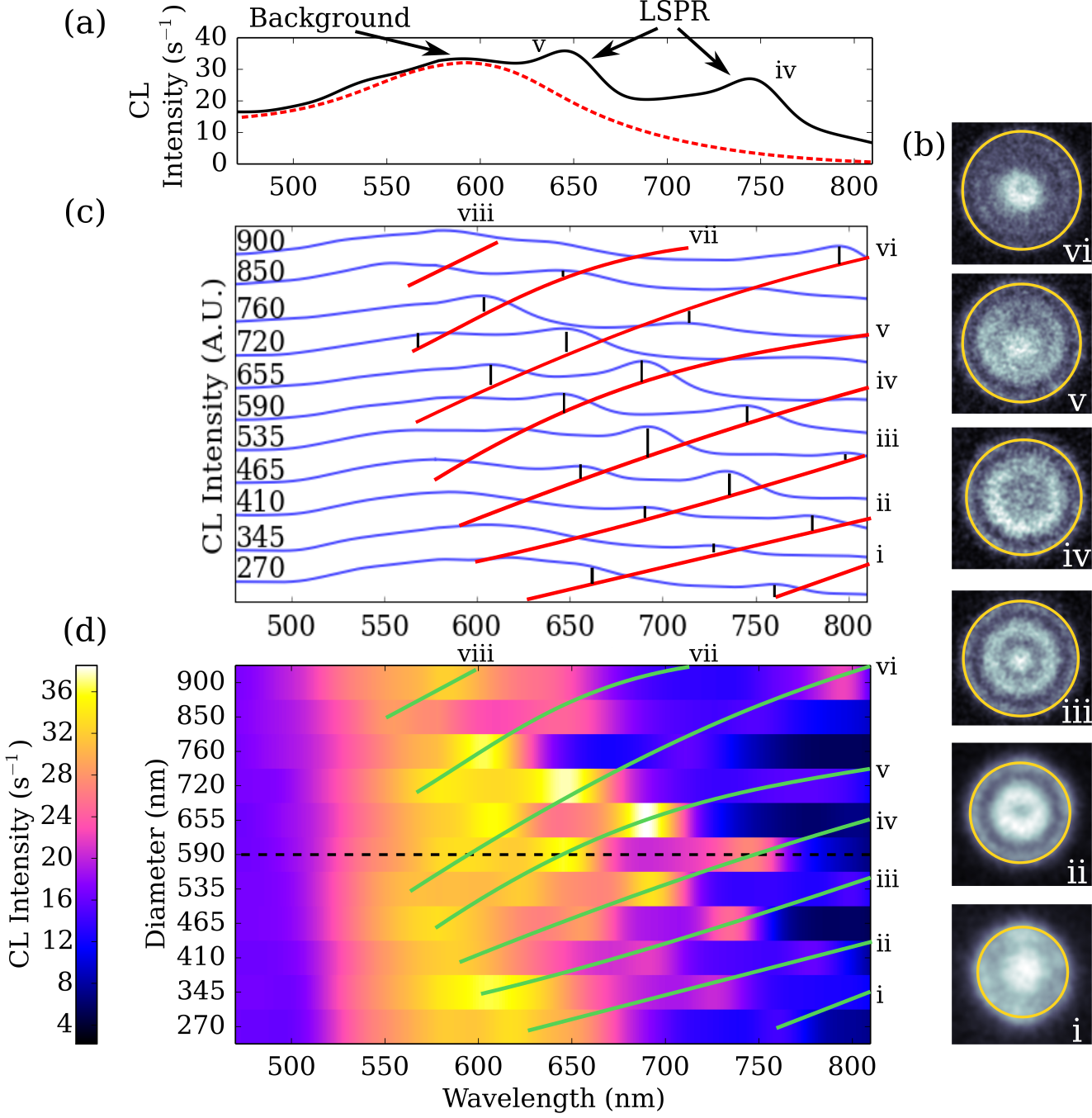}
\caption{(a) Typical CL spectrum of a Au patch plasmonic antenna of
	\unit{590}{\nano\meter} diameter. The background contribution is shown in red
	dashed line. The LSPRs of the antenna are indicated, with the Roman number
	corresponding to the LDOS images in panel (b) and branches in panel (d). 
	(b) Corresponding energy-filtered CL images revealing
	the LDOS pattern associated with the first six resonances branches in (c). 
	The yellow circles represent the antenna physical size.
	(c) Waterfall plot of the CL spectra obtained for different antenna diameters. 
	The red lines follow the position of the resonances (marked in black) and serve 
	as guides for the eye. They are numbered from the lowest order (smallest antenna) 
	to the largest order in small Roman numbers.
	(d) Corresponding intensity color-coded spectral map of the antennas CL signal 
	as a function of wavelength and antenna diameter. The resonance branches are evidenced 
	in green lines as guides for the eye, with the same Roman numbering.
	Note that we have fixed the width of each spectrum,
	resulting in an unevenly spaced vertical scale, to better show each resonance
	position. The black dashed line corresponds to the spectrum in (a).}
\label{fig:SpectralMapLDOS}
\end{figure}

Let us first describe the spectral properties of the Au patch antennas. To
obtain the antennas CL spectrum, we raster scan the electron beam over the
antenna surface, using a square scanning area exactly circumscribing the disk.
The beam scanning time is set to be much smaller than the integration time
($\sim$\unit{1}{\minute}) so that every point on the antenna surface can be
considered as equally excited. A background spectrum is acquired by scanning the
exact same area on a region where no antenna is present, recording the CL
response of the bare substrate. This substrate spectrum is subtracted from the
antenna's CL response. We note that the background luminescence from the
substrate has an intensity of $\sim\unit{4000}{\second}^{-1}$, while the CL
signal originating from localized surface plasmon modes supported by the antenna
is $\sim\unit{20}{\second}^{-1}$ in intensity. The background subtraction
procedure does not suppress entirely the contribution of the substrate
luminescence to the collected CL signal. A typical antenna spectrum is presented
in Fig.~\ref{fig:SpectralMapLDOS}(a) (\unit{590}{\nano\meter} diameter). It presents a broad peak  indicated in red
dashes. This signal comes from residual luminescence due to Au interband
transitions for wavelengths below \unit{600}{\nano\meter}\cite{Beversluis2003} as
well as from the diffraction by the patch edge of the surface plasmon polaritons
excited outside the antenna region at larger wavelength.

Sharper peaks are visible at 650 and \unit{750}{\nano\meter}, which we attribute
to localized surface plasmon resonances (LSPR) of the antenna. 
CL images of the LDOS corresponding to each LSPR are obtained by
slowly scanning the electron beam over the antenna surface, collecting the light
emitted in a \unit{50}{\nano\meter} spectral bandwidth around the resonance.
This filtered CL emission is then detected by an APD and correlated to the
electron beam position. The spectral integration bandwidth is selected to
maximise the collected intensity, matching the resonance bandwidth but ensuring
that we collect only light emitted from a single resonance. In the rest of the
article, the LDOS images are normalized in intensity. The typical emission
probability is of the order of $10^{-6}$ photons per incident electrons. More
information on the LDOS image acquisition is given in the Appendix section. Note
that the CL images do not reveal the antenna mode structure, but rather map the
probability of exciting a given antenna mode at each electron beam
position.
The LDOS images corresponding to the two LSPR peaks observed in Fig.~\ref{fig:SpectralMapLDOS}(a)
are shown in the corresponding iv and v insets in Fig.~\ref{fig:SpectralMapLDOS}(b). 

We subsequently repeat this procedure for several antenna diameters, 
and obtain the antenna spectrum with its LSPR peaks and their corresponding LDOS patterns. 
In Fig.~\ref{fig:SpectralMapLDOS}(c), we gather all the antenna spectra in the form of 
a waterfall plot. The experimentally measured antenna diameters are indicated, and fabrication imperfections 
are responsible for the non-even spacing between the different antennas sizes. 
It is also represented by the color-coded intensity spectral map 
in Fig.~\ref{fig:SpectralMapLDOS}(c) which represents the  
CL spectra as a function of wavelength and increasing diameter. 
Note that the vertical scale has a fixed width for each spectrum. 
This representation will be used again in Fig.~\ref{fig:SpectralMapVsMaterial}. 

The LDOS imaging allows us to track a given LSPR with respect to the antenna diameter change. 
It reveals that the successive LSPRs of the patch red-shift with increasing antenna diameter. 
The red-shifting resonances branches are marked with red solid lines serving as guides for the eye. 
All the resonances belonging to a given branch show the same spatial structure, 
as represented in Fig.~\ref{fig:SpectralMapLDOS}(b) for the first six branches. 
Each pattern is composed of one or more concentric rings. 
In addition, considering successive LDOS patterns (from i to vi)
we observe an alternation of bright and dark antenna center, in agreement with 
previous reports \cite{Mohtashami2014}. 
Furthermore, we do not observe any azimuthal dependence of the LDOS. 
This will be discussed in the modelling section. 
\\

\section{Semi-analytical model}

\begin{figure}[htbp]
\centering\includegraphics[width=\linewidth]{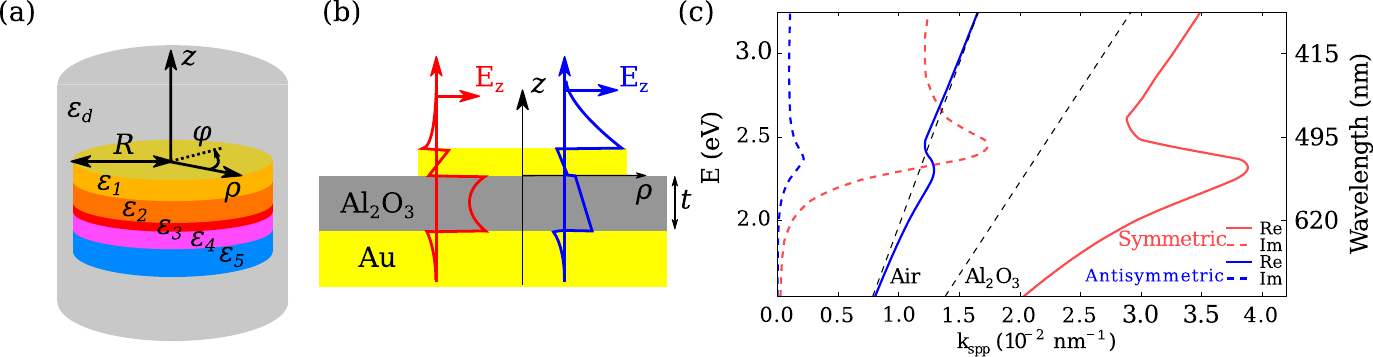}
\caption{(a) Sketch of the geometry considered in the analytical model developed
	in \cite{Mohtashami2014}. (b) Sketch of the experimental geometry, with the
	two vertical mode profiles sustained by the patch antenna: symmetric (red) and
	antisymmetric (blue). Note that in our case, only the patch is of finite radius
	$R$. (c) Dispersion relations of the symmetric (red) and antisymmetric (blue)
	mode profiles. The real part of the wave vector $k_{n,m}$ is represented in
	solid lines, and its imaginary part in dashed lines. The black dashed lines are
	the photonic dispersion relations in air and alumina. }
\label{fig:TheoryPresentation}
\end{figure}

The optical properties of plasmonic patch antennas can be described by the
semi-analytical model of ref.~\cite{Mohtashami2014} which is adapted from the
fully analytical work of ref.~\cite{Filter2012}. The analytical model considers
an arbitrary stack of circularly symmetric layers of fixed radius $R$, different
thickness $t_j$ and permittivity $\varepsilon_j$, where $j$ is the layer number,
embedded in an homogeneous medium of permittivity $\varepsilon_d$. A sketch of
the geometry is shown in Fig.~\ref{fig:TheoryPresentation}(a). The stack is
considered as a circular cavity supporting Bessel-type surface plasmon polariton
(SPP) modes. The dispersion relation of the different SPP modes and their
vertical mode profile are numerically calculated from the equivalent infinite
($R \rightarrow \infty$) multi-layer system~\cite{Davis2009}. In the case of our
insulator-metal-insulator-metal (IMIM) geometry, only two distinct propagating
SPP modes are present, referred to as \emph{symmetric} and \emph{antisymmetric}
due to the symmetry of the $E_z$ component of the electric field. The vertical
profiles of $E_z$ are sketched in Fig.~\ref{fig:TheoryPresentation}(b) in red
(symmetric mode) and blue (antisymmetric mode). The symmetric mode has most of
its energy confined inside the spacing layer, while the antisymmetric mode is
rather confined at the top air/metal interface. The respective dispersion
relations of the two modes are presented in Fig~\ref{fig:TheoryPresentation}(c),
where the solid lines represent the real part of the dispersion relation, and
the dashed lines its imaginary part. The antisymmetric mode shows low losses but
is almost index-matched with the free-space dispersion relation in air, and is
thus poorly confined at the metal surface. The symmetric mode significantly
deviates from the free-space propagation in alumina, indicating a strong
confinement. It also suffers from greater losses than the antisymmetric one.
Because it presents a very strong electric field inside the alumina layer and
matches the polarisation of the exciting field created by the electrons, this
mode plays a dominant role in the CL experiment. Hence, as in
refs.~\cite{Kuttge2010,Mohtashami2014}, we restrict the subsequent analysis to
the symmetric vertical mode profile.

The electric field is decomposed into eigenmodes of the circular 
geometry in which the $z$, $\rho$ and $\varphi$ dependences can be separated as:
\begin{equation}
E^{n,m}_z\left(\rho,z\right) = a(z)J_m\left(k_{n,m} \rho\right)\exp\left(im\varphi\right)
\label{eq:Eigenfield}
\end{equation}
where $a(z)$ is the vertical mode profile sketched in red in
Fig.~\ref{fig:TheoryPresentation}(b), $J_m$ is the Bessel function of the first
kind of azimuthal order $m$, $\rho$ is the radial vector and $k_{n,m}$ is the
surface plasmon polariton wave vector supported by the infinite multilayer
structure following the dispersion relation of
Fig.~\ref{fig:TheoryPresentation}(c). The eigenfunctions are further determined
by the resonance condition:
\begin{equation}
\mathrm{Re}\left[k_{n,m}\right]2R + \varphi_m = 2x_n\left(J_m\right)
\label{eq:ResonanceCondition}
\end{equation}
where $\mathrm{Re}\left[k_{n,m}\right]$ denotes the real part of the SPP wave
vector. The phase factor $\varphi_m$ accounts for the phase shift acquired upon
reflection at the patch edge $\rho=R$. $x_n$ is the $n-th$ zero of the Bessel
function $J_m$. While $\varphi_m$ can be numerically calculated for each 
azimuthal number $m$ according to ref.\cite{Filter2012}, 
we used the empirical formulas and method reported in
ref.~\cite{Mohtashami2014} instead. The main approximation is that $\varphi_m$
is considered the same for all antenna modes $(n,m)$. 
Equation \ref{eq:ResonanceCondition} gives the resonant frequencies
$\omega_{n,m}$ of the disk after inversion of the SPP dispersion relation. The
LDOS $\Gamma\left(\omega, \rho\right)$ probed during the CL experiment can thus
be expressed as:
\begin{equation}
\Gamma\left(\omega,\rho\right) \propto \sum_{n,m} \left|E^{n,m}_z(\rho) \right|^2 
						\mathcal{L} \left(\omega,\omega_{n,m},\gamma_{n,m} \right)
\label{eq:LDOS}
\end{equation}
where the energy of the mode profile $E^{n,m}_z(\rho)$ is normalized in space through 
$\iint 2\pi \rho \left|E^{n,m}_z(\rho) \right|^2 \mathrm{d}\rho \mathrm{d}z  = 1$, 
and $\mathcal{L}\left(\omega,\omega_{n,m},\gamma_{n,m}\right)$ is a normalized
Lorentzian function centred at frequency $\omega_{n,m}$ with a full width at
half maximum $\gamma_{n,m}$ so that $\int
\mathcal{L}\left(\omega,\omega_{n,m},\gamma_{n,m}\right) \mathrm{d}\omega = 1$.
Following ref.~\cite{Mohtashami2014}, the width of the Lorentzian function is
given by:
\begin{equation} 
\gamma_{n,m}/\omega=(2\mathrm{Im}\left[k_{n,m}\right])^2+\gamma_0(1+\omega_{\mathrm{norm}}^b)^{-1}
\end{equation} 
where $\gamma_0$ and $b$ are adjustable parameters used to fit the resonance
bandwidth of the antennas, $\mathrm{Im}\left[k_{n,m}\right]$ is the imaginary
part of the SPP wave vector, and $\omega_{\mathrm{norm}} =
\omega/(\unit{1800}{\tera\hertz})$. The frequency broadening of the antennas
resonances describes the loss-induced frequency broadening of the plasmonic
patches. Note that as the LDOS is related to the square modulus of the electric
field, one can only detect the radial dependence of the plasmon mode, and not
its azimuthal dependence.

\begin{figure}[htbp]
\centering\includegraphics[width=\linewidth]{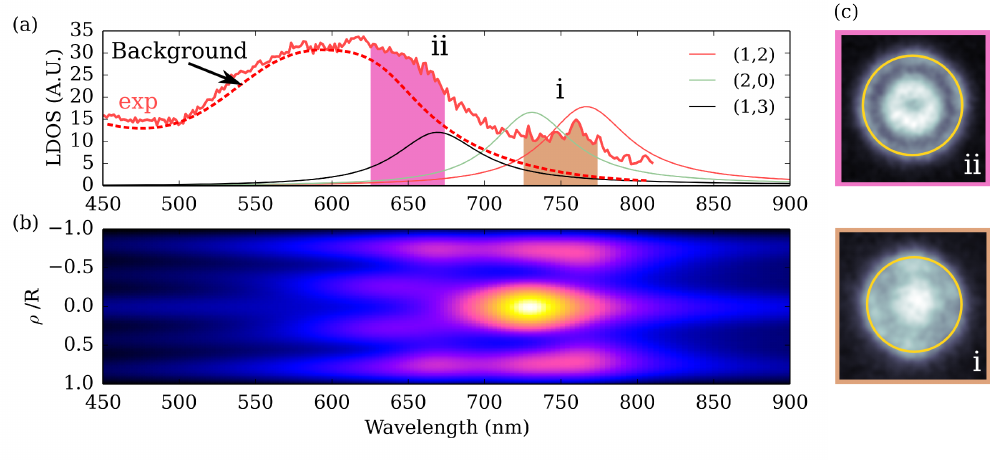}
\caption{(a) Comparison between the measured antenna CL spectrum (thick red
	line) and the simulated resonances (thin lines) of a \unit{270}{\nano\meter}
	diameter antenna. The background contribution is indicated in red dashes. The
	principal Bessel contributions to the simulated spectrum are shown in thin solid
	lines. The corresponding Bessel numbers $(n,m)$ are indicated in the legend. (b)
	Simulated spatial LDOS map: as it has no azimuthal dependence, it is represented
	as the function of the wavelength and the normalized radial coordinate $\rho/R$
	which varies form -1 to +1 along the diameter of the antenna. Note the
	separation between the $m=0$ mode producing a maximum of LDOS at the antenna
	center, and the $m \neq 0$ modes. (c) Measured CL images using the integration
	bandwidth indicated by the colored area in (a). The yellow circles indicate the
	antenna physical size. The Roman numbers refer to the mode branches numbers in
	Figs.~\ref{fig:SpectralMapLDOS}(b)--\ref{fig:SpectralMapLDOS}(c).
}
\label{fig:Comparison141006}
\end{figure}

Using the previous relations we compute the different resonances contributing to
the LDOS for each radius of the the fabricated antennas. The fitting parameter
$\varphi_m$ was obtained in \cite{Mohtashami2014} by matching the resonance
of the fundamental antenna mode to the results of numerical simulations.
However, due to the large disk size, this mode had a very low energy and could
not be experimentally observed. Here instead, we match the position of all the
antenna modes with the experimentally determined resonances of branches i, ii
and iii in Fig.~\ref{fig:SpectralMapLDOS}(c). In this case, we find that a
correction on the SPP dispersion relation is necessary to reproduce the
experimental mode dispersion. The semi-analytical model thus seem to
underestimates the mode wave vector by a factor 1.5 to 2. This might be due to
the fact that the model considers the disks as cavities for propagating SPPs,
whereas for small disks the large lateral confinement may result in larger SPP
effective wave vectors, as already described in the case of plasmon waveguides
\cite{Berini2000,Berini2001}. Fitting the experimental data, we find
$\gamma_0=0.05$, $b=1.7$, and $\varphi_m=0.5$.

Comparing the results of the simulations with the experimentally determined
spectra and LDOS reveals further information on the mode structure supported by
the patch antennas. Figure \ref{fig:Comparison141006}(a) shows the comparison
between the measured spectrum (thick red line) and the predicted resonances
(thin lines) for a \unit{270}{\nano\meter} diameter antenna (smallest antenna in
Fig.~\ref{fig:SpectralMapLDOS}(c)). The main contribution to the LDOS originates
from the three Bessel modes indicated in thin solid lines, with their radial and
azimuthal number $(n,m)$ given in the legend. The simulated spatial LDOS map is
shown in Fig.~\ref{fig:Comparison141006}(b), while the top insets in panel (a)
show the measured LDOS using a spectral bandwidth represented by the shaded area
in the spectrum. We can see that the resonance at \unit{750}{\nano\meter} is
composed of the $(n=1,m=2)$ and $(n=2,m=0)$ modes. The $m=0$ modes are the only
ones contributing to the LDOS at the antenna center, for symmetry reasons. Note
that the $m=0$ mode (green) appears at a slightly shorter wavelength than the
$(n=1,m=2)$ mode (red). On the other hand, the resonance at
\unit{650}{\nano\meter} is composed of a single $(n=1,m=3)$ Bessel mode.
Comparing the spatial LDOS maps with the experimental LDOS images allows us to
confirm the fitting parameter.

As noted in \cite{Mohtashami2014}, it is important to realize that the
patch antenna resonances are composed of superposition of several Bessel modes
with increasing quantum numbers. This superposition arises from the overlap
between the frequency broadened Bessel modes due to ohmic and radiation losses
\cite{Mohtashami2014}. However, in our case, the small size of our patch antenna
allows us to image LDOS corresponding to single mode resonances. Doing so, we
arrive at a one-to-one correspondence between the LDOS model and the
experimental CL images, which was not possible in \cite{Mohtashami2014}.
Note that the relative amplitude between the resonances is not reproduced. This
is because the model only describes the LDOS supported by the particle, while CL
is sensitive to the radiative component of the LDOS. The radiative efficiency
and radiation pattern of each resonance is thus left out in the model and is
responsible for this discrepancy. Finally, the broad continuum of CL signal
below \unit{600}{\nano\meter} is not reproduced by the model, because it is
simply due to luminescence from the Au layers and thus not related to the
plasmonic mode structure of the antennas.

\section{Towards lower orders}

Figure \ref{fig:Comparison141006} shows that the antenna can sustain resonances
corresponding to a single plasmon mode. 
Reaching low order mode operation, or even single mode operation is important to
efficiently couple the patch antennas to localized emitters placed inside the
spacing layer, such as in \cite{Belacel2013}. Large patch antennas support
a high number of competing mode, which gives them their interesting directional
beaming properties, as shown in depth in \cite{Mohtashami2014}. However, a
large radiative enhancement requires a strong coupling between the emitter and a
single radiative mode of the antenna. This is why the strongest reported Purcell
factors in the patch geometry involve small metallic particles as in
refs.~\cite{Hoang2016,Chikkaraddy2016}. Hence, a trade-off has to be found
between directionality and radiative rate enhancement when coupling
nano-emitters to such plasmonic patch antennas.

\begin{figure}[htbp]
\centering\includegraphics[width=\linewidth]{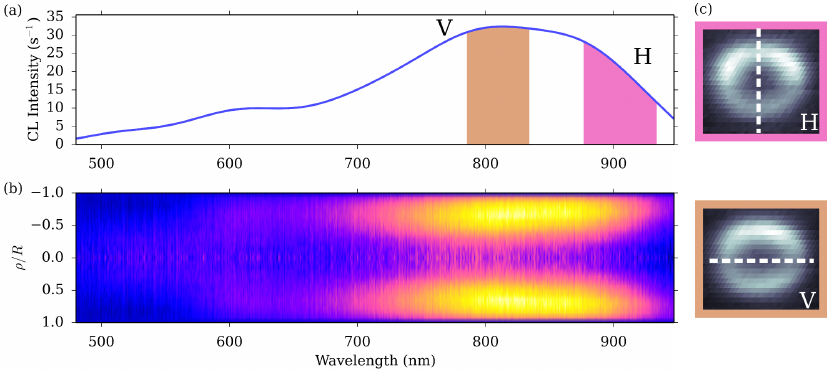}
\caption{CL spectrum (a) and hyperspectral CL map (b) of a
	\unit{170}{\nano\meter} diameter antenna. The hyperspectral map in (b) is
	obtained by summing the spectra of all the pixels whose distance from the
	antenna center is between $\rho$ and $\rho+d\rho$ and normalized to the
	corresponding surface element. It is symmetric by construction with respect to
	$\rho=0$. It clearly reveals the fundamental antenna mode $n=m=1$. (c) LDOS CL
	image for the spectral integration bandwidths H and V indicated in (a). The
	degeneracy of the mode is lifted due to the imperfect shape of the disk and lead
	to the breaking of the circular symmetry. As a consequence we observe a
	horizontal (H) and vertical (V) splitting of the corresponding LDOS images.}
\label{fig:LDOSHyperspectral}
\end{figure}

The lowest order radiative mode that can
be sustained by the patch antenna is the fundamental $(n=1,m=1)$ dipole-like
mode. To reveal this fundamental mode, we fabricated smaller Au antennas on an
Al mirror. The use of the Al mirror strongly reduces the luminescence
contribution from the substrate, as will be evidenced in the next section. The
CL spectrum and hyperspectral map of the CL signal as a function of the
wavelength and antenna reduced radius are presented in
Fig.~\ref{fig:LDOSHyperspectral} for a \unit{170}{\nano\meter} diameter antenna.

This antenna clearly exhibits a dipole-like LDOS spatial pattern, characteristic
of the fundamental $(n=1,m=1)$ antenna mode, as can be seen in
Fig.~\ref{fig:LDOSHyperspectral}(b). However, the spectrum presented in
Fig.~\ref{fig:LDOSHyperspectral}(a) exhibit a very broad unresolved double peak
resonance centred at \unit{845}{\nano\meter}. Using the hyperspectral mode of
the setup, we image the LDOS supported by the antenna on the high and low energy
side of this resonance. Fig.~\ref{fig:LDOSHyperspectral}(c) shows the LDOS using
the integration bandwidths coloured in Fig.~\ref{fig:LDOSHyperspectral}(a). It
reveals that due to fabrication imperfections, the antenna in not perfectly
circular, resulting on an apparent splitting of the fundamental $(n=1,m=1)$ mode
and breaking of the circular symmetry. We note the the CL signal seems stronger
on the upper half of the antenna for both the horizontal (H) and vertical (V)
modes in Fig.~\ref{fig:LDOSHyperspectral}(c). We attribute this effect to the
progressive SEM induced contamination of the antenna surface as we scan the beam
over it (total acquisition time is $\sim$\unit{30}{\minute})

\section{Towards higher energies}

\begin{figure}[htbp]
\centering\includegraphics[width=\linewidth]{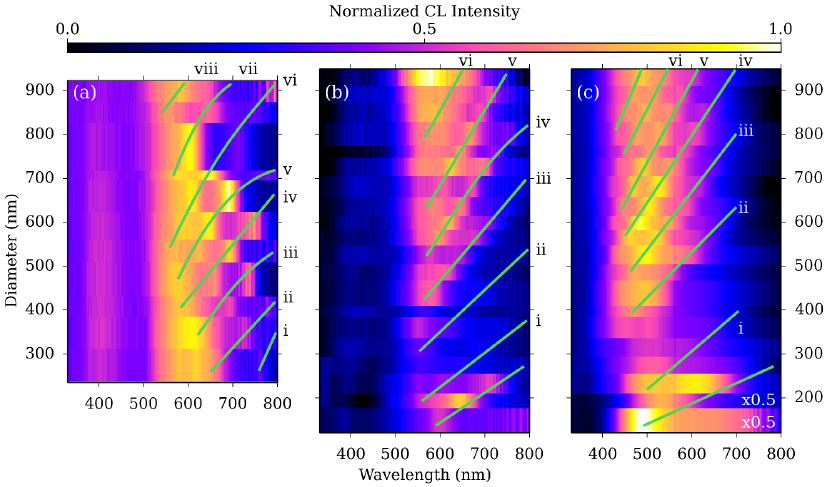}
\caption{CL intensity spectral maps of Au patch antennas on a Au mirror (a), Au
	patch antennas on an Al mirror (b) and Al patch antennas on an Al mirror (c)
	representing the CL spectrum of each antenna as a function of wavelength and
	diameter. Note that in panel (c), the first two spectra show stronger
	resonances, and have been divided by 2 to increase the overall contrast. Panel
	(a) also corresponds to Fig.~\ref{fig:SpectralMapLDOS}(c). }
\label{fig:SpectralMapVsMaterial}
\end{figure}

Having completely characterize the mode structure of small circular Au plasmonic
patch antennas, we fabricated Al patch antennas on an Al mirror to extend their
operation wavelength to higher energies.  Figure \ref{fig:SpectralMapVsMaterial}
gathers the spectral information on patch antennas as a function of patch
diameter and wavelength. Panels (a) to (c) correspond to the three different
kinds of samples presented in Fig.~\ref{fig:PatchsPresentation}, respectively Au
patch on Au film, Au patch on Al film and finally Al patch on Al film. As
mentioned previously, the comparison between
Fig~\ref{fig:SpectralMapVsMaterial}(a) and (b) show that the presence of the Al
mirror strongly reduces the luminescence contribution from the substrate,
resulting in a low background signal for wavelengths below
\unit{550}{\nano\meter} in panel (b). Further comparison between panels (b) and
(c) show the dispersion relation of the successive antenna resonances, including
the dipole-like, fundamental mode for small diameters. As seen from
Fig.~\ref{fig:SpectralMapVsMaterial}(b), the Au patch antennas on an Al mirror
provide optical resonances for wavelengths from 550 to \unit{750}{\nano\meter},
while the Al patch antennas in Fig.~\ref{fig:SpectralMapVsMaterial}(c) exhibit
resonances spanning wavelengths from \unit{450}{\nano\meter} to
\unit{650}{\nano\meter}. We note however that there is a trade-off between the
vertical confinement of the mode in the oxide layer and the high energy
operation of the patch structures. Because most of the energy of the symmetric
vertical mode profile is confined inside the oxide layer, increasing the
strength of the mode requires a smaller oxide thickness. This leads in return to
a stronger confinement of the electric field and hence a stronger refractive
index sensed by the mode, which results in a red-shift of the resonant
wavelengths for a given patch diameter. Furthermore, the plasmonic response of
Al particles is limited to the blue and green parts of the visible
electromagnetic spectrum, since a broad interband transition is present in Al at
\unit{800}{\nano\meter} \cite{Gerard2014}.

LDOS imaging experiment on each resonance confirm that the change
in material allows to tune the resonant wavelengths of the antennas but does not
affect their supported mode structure, which is only characteristic of the
antenna geometry. We note that due to the roughness of the Al film, LDOS imaging
was more complicated in the case of Al antennas. The presence of such large Al
grains produces a lot of parasitic signal arising from surface plasmon modes
strongly localized on the grains. Parasitic CL signal also originates from the
diffraction by the surface roughness of the continuum of surface plasmon
polaritons excited by the electron beam.

\section{Conclusion}

In conclusion, we have used cathodoluminescence spectroscopy and imaging to
investigate the spectral and spatial plasmonic properties of circular plasmonic
patch antennas made of Au and Al. In a first step, we have characterized Au
patch antennas on an Au mirror. Their characteristic spectrum shows several
resonances above \unit{600}{\nano\meter}. Energy-resolved CL imaging allowed us
to image the radiative LDOS corresponding to the resonances. We have linked the
CL properties of these antennas to a semi-analytical model presented in
ref.~\cite{Mohtashami2014}. Contrary to what is reported in this reference, we
arrive here at a one-to-one correspondence between the measured and simulated
LDOS at the expense of correcting the surface plasmon wave vector dispersion
relation by an empirical factor. The increasing complexity of the mode structure
supported by the antennas comes from the broadening of the modes due to ohmic
and radiation losses. Hence, a single antenna resonance can be composed of
several overlapping antenna modes. We then further reduced the size of the
circular antennas and used hyperspectral CL imaging to reveal the fundamental
mode of Au patch antennas on an Al mirror. Finally, using Al patch antennas on
an Al mirror, we demonstrated the possibility to fabricate Al patch antennas
providing optical resonances down to \unit{450}{\nano\meter}.

\appendix

\section{CL System}
\begin{figure}[htbp]
\centering\includegraphics[width=\linewidth]{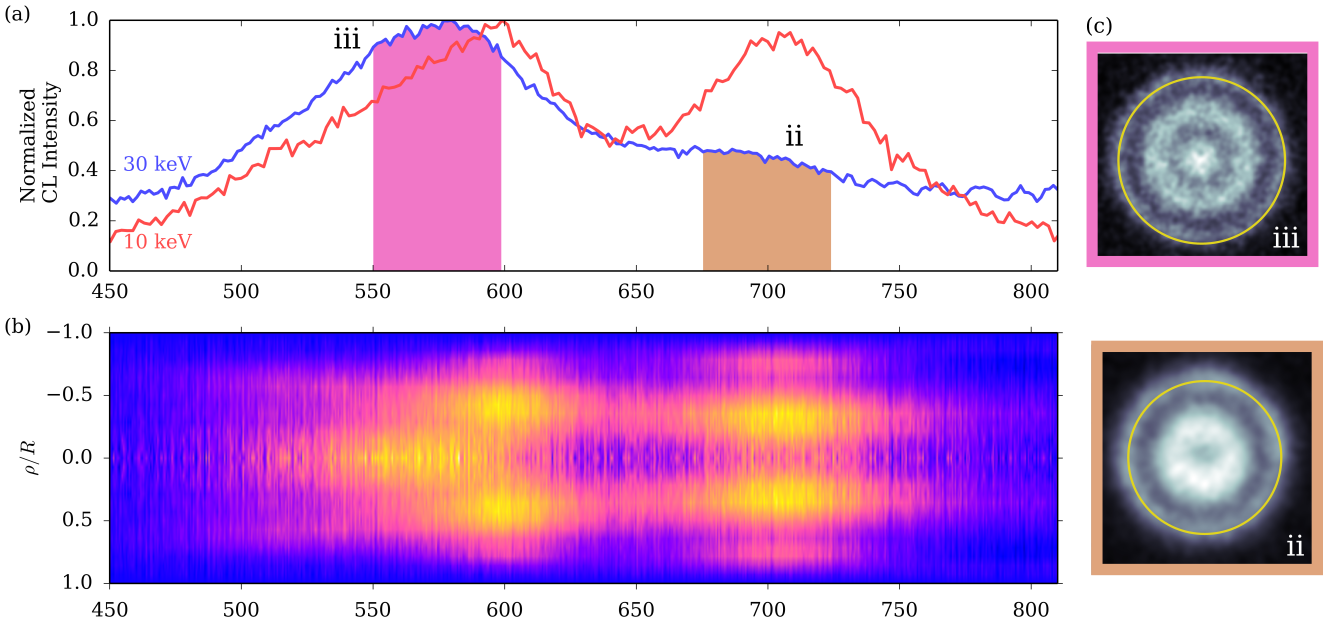}
\caption{(a) CL spectra for a \unit{435}{\nano\meter} diameter Au antenna on an
	Al mirror, as presented in Fig~\ref{fig:SpectralMapVsMaterial}(b). The red line
	corresponds to the \unit{10}{\kilo\electronvolt} excitation, while the blue line
	corresponds to the \unit{30}{\kilo\electronvolt} excitation. (b) Hyperspectral
	LDOS imaging with the \unit{10}{\kilo\electronvolt} beam as a function of
	wavelength and reduced antenna radius. (c) LDOS images obtained using the
	\unit{30}{\kilo\electronvolt} setup by integrating the CL signal in the shaded
	spectral regions. They correspond to the second and third mode branches in
	Fig.~\ref{fig:SpectralMapLDOS}(b)--(c), starting from the smallest antenna. The
	yellow circles indicate the antenna physical size.
}
\label{fig:CompareCLSystems}
\end{figure}

To perform LDOS imaging, the CL signal in filtered in energy using the exit slit
of the spectrometer and sent on an APD. The APD output is connected to an
electronic pulse generator. We ensure that the amplitude and time of the pulse
are set so that we operate in a photon counting mode. Typically, the pulse
duration is one third of the pause time of the electron beam on each position.
The resulting image is thus composed of pixels of discrete intensity values. A
digitalisation step is perform to convert these discrete intensity values into a
number of counting events. To account for the spatial extension of the electron
beam, a spatial Gaussian filter of \unit{10}{\nano\meter} width is applied on
the signal.

We can compare the results obtained from the two CL setups using
Fig.~\ref{fig:CompareCLSystems}, where the CL properties of a
\unit{435}{\nano\meter} diameter Au antenna on an Al mirror are shown. Panel (a)
represents the two CL spectra and panel (b) the hyperspectral LDOS map obtained
with the \unit{10}{\kilo\electronvolt} beam. Panel (c) shows the LDOS images
obtained with the \unit{30}{\kilo\electronvolt} beam and integrated over the
spectral width shaded in under the spectrum. Comparing the LDOS obtained using
hyperspectral imaging in Fig.~\ref{fig:CompareCLSystems}(b) and
Fig.~\ref{fig:CompareCLSystems}(c) show the same mode structure, i.e. the bright
center and double ring structure of the resonance at \unit{575}{\nano\meter} and
the double ring structure of the resonance at \unit{700}{\nano\meter}.

\section*{Funding}
This work was supported by the French National Research Agency, 
Labex LANEF du Programme d'Investissements d'Avenir ANR-10-LABX-51-01 through the LANEF PhD Grant and the 
LANEF equipment project TREE.

\section*{Acknowledgments}
We acknowledge the help of Institut N\'eel technical support teams Nanofab
(clean room) and  optical engineering (CL, Fabrice Donatini). The hyperspectral CL experiments 
were performed on the Nanocharacterization platform (PFNC) at MINATEC.

\end{document}